\documentclass[sigconf,authorversion=true,]{acmart}
\usepackage{subcaption}

\AtBeginDocument{%
  \providecommand\BibTeX{{%
    \normalfont B\kern-0.5em{\scshape i\kern-0.25em b}\kern-0.8em\TeX}}}

\setcopyright{rightsretained}
\copyrightyear{2022}
\acmYear{2022}
\acmConference{SIGGRAPH '22 Emerging Technologies }{August 07-11, 
2022}{Vancouver, BC, Canada}\acmBooktitle{Special Interest Group on Computer 
Graphics and Interactive Techniques Conference Emerging Technologies (SIGGRAPH 
'22 Emerging Technologies ), August 07-11, 2022}\acmDOI{10.1145/3532721.3546117}
\acmISBN{978-1-4503-9363-8/22/08}

\begin{document}

\title{Meta Avatar Robot Cafe: Linking Physical and Virtual Cybernetic Avatars to Provide Physical Augmentation for People with Disabilities}

\author{Yoichi Yamazaki}
\email{yamazaki@he.kanagawa-it.ac.jp}
\orcid{0000-0001-6402-4587}
\affiliation{%
  \institution{Kanagawa Institute of Technology}
  \city{Atsugi}
  \state{Kanagawa}
  \country{Japan}
}

\author{Tsukuto Yamada}
\email{s1933054@cco.kanagawa-it.ac.jp}
\affiliation{%
  \institution{Kanagawa Institute of Technology}
  \city{Atsugi}
  \state{Kanagawa}
  \country{Japan}
}

\author{Hiroki Nomura}
\email{s1933010@cco.kanagawa-it.ac.jp}
\affiliation{%
  \institution{Kanagawa Institute of Technology}
  \city{Atsugi}
  \state{Kanagawa}
  \country{Japan}
}

\author{Nobuaki Hosoda}
\email{s1933040@cco.kanagawa-it.ac.jp}
\affiliation{%
  \institution{Kanagawa Institute of Technology}
  \city{Atsugi}
  \state{Kanagawa}
  \country{Japan}
}

\author{Ryoma Kawamura}
\email{s1833019@cco.kanagawa-it.ac.jp}
\affiliation{%
  \institution{Kanagawa Institute of Technology}
  \city{Atsugi}
  \state{Kanagawa}
  \country{Japan}
}

\author{Kazuaki Takeuchi}
\email{k.takeuchi@orylab.com}
\affiliation{%
  \institution{OryLab Inc.}
  \city{Chuo-ku}
  \state{Tokyo}
  \country{Japan}
}

\author{Hiroaki Kato}
\email{h.kato@orylab.com}
\affiliation{%
  \institution{OryLab Inc.}
  \city{Chuo-ku}
  \state{Tokyo}
  \country{Japan}
}

\author{Ryuma Niiyama}
\email{niiyama@meiji.ac.jp}
\orcid{0000-0002-9072-8251}
\affiliation{%
  \institution{Meiji University }
  \city{Kawasaki}
  \state{Kanagawa}
  \country{Japan}
}

\author{Kentaro Yoshifuji}
\email{ory@orylab.com}
\affiliation{%
  \institution{OryLab Inc.}
  \city{Chuo-ku}
  \state{Tokyo}
  \country{Japan}
}

\renewcommand{\shortauthors}{Yamazaki and Yamada, et al.}

\begin{abstract}
Meta avatar robot cafe is a cafe that fuses cyberspace and physical space to create new encounters with people. We create a place where people with disabilities who have difficulty going out can freely switch between their physical bodies and virtual bodies, and communicate their presence and warmth to each other.
\end{abstract}



\begin{teaserfigure}
 \begin{minipage}{0.31\linewidth}
  \centering
  \includegraphics[height=35mm]{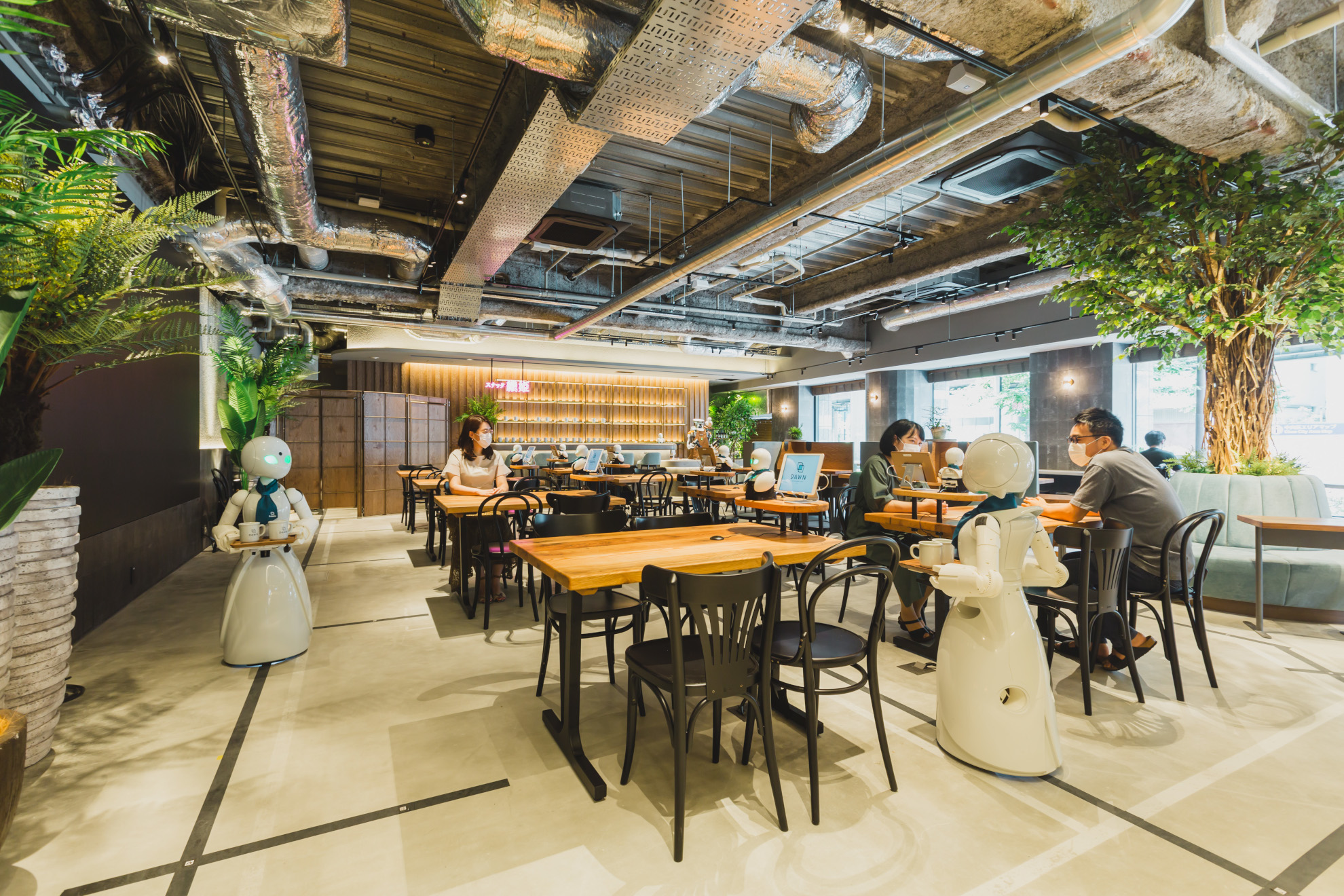}
  \subcaption{Avatar robot cafe DAWN ver. $\beta$.}
  \label{fig:one}
 \end{minipage}
 \begin{minipage}{0.32\linewidth}
  \centering
  \includegraphics[height=35mm]{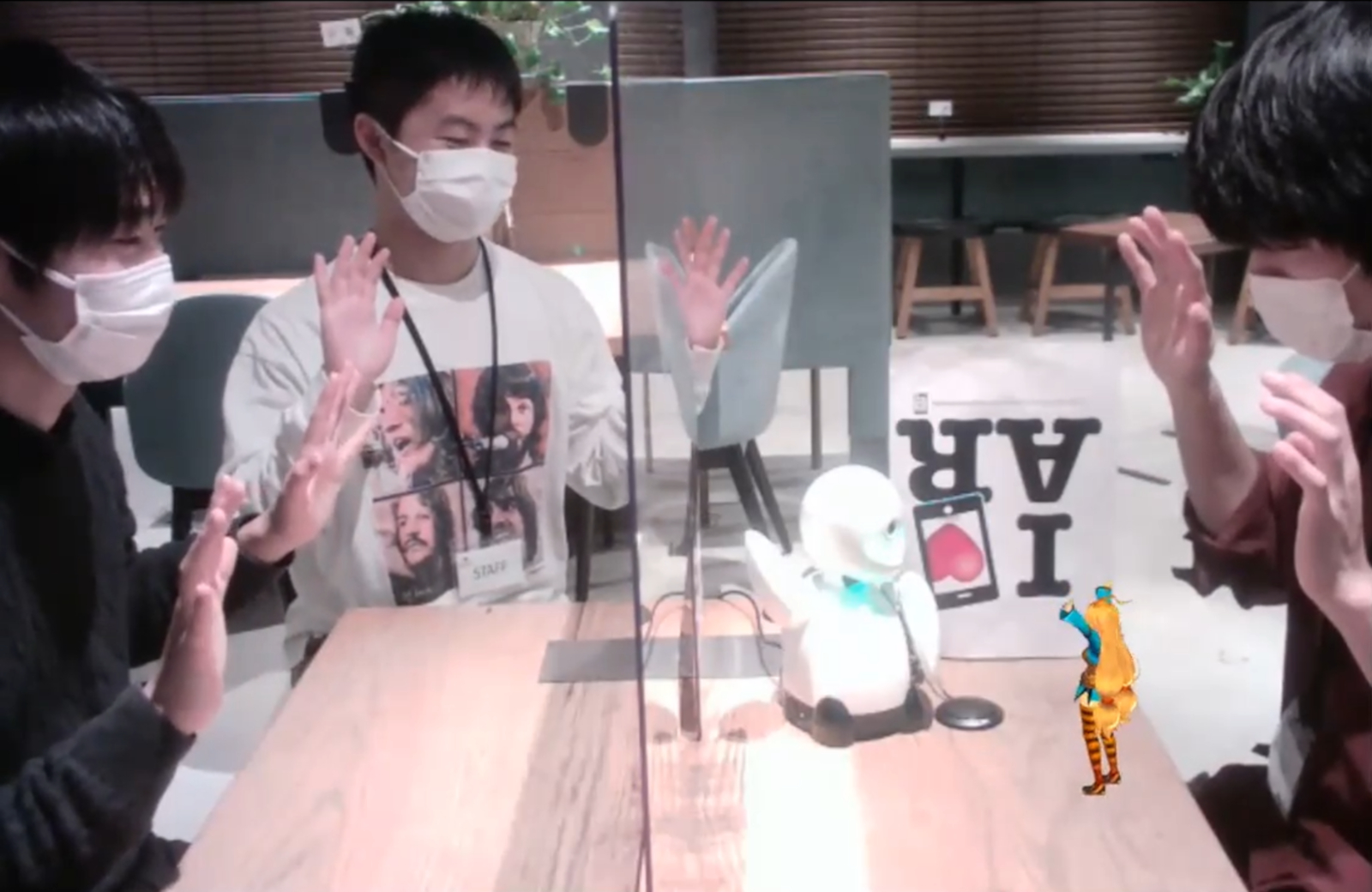}
  \subcaption{AR operation (\textcopyright UTJ/UCL)}
  \label{fig:two}
 \end{minipage}
 \begin{minipage}{0.36\linewidth}
  \centering
  \includegraphics[height=35mm]{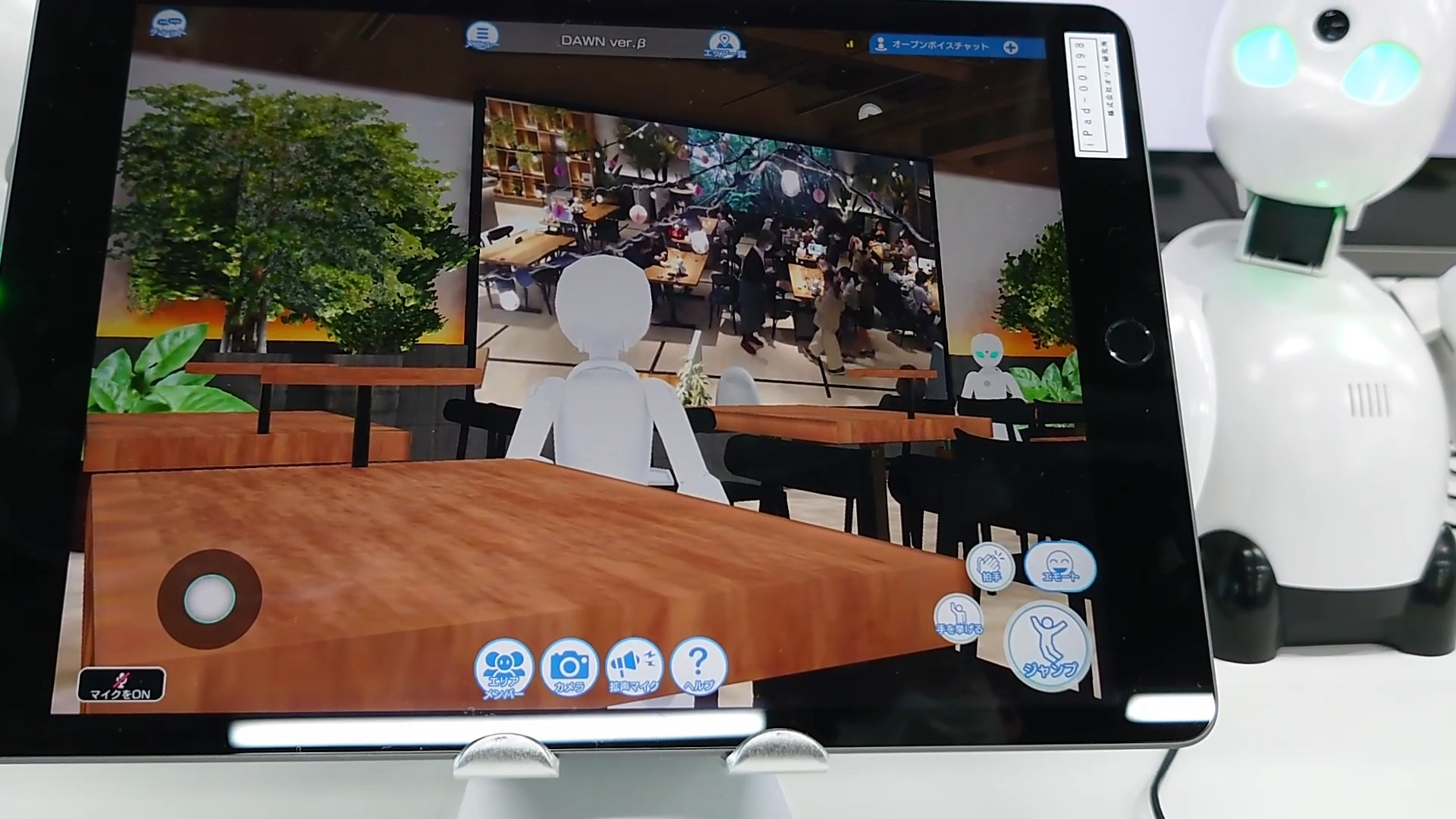}
  \subcaption{VR operation}
  \label{fig:three}
 \end{minipage}
 \caption{Operation screens by a pilot who has disabilities at the avatar robot cafe DAWN ver. $\beta$.}
 \Description{Customer service scenes at Avatar robot cafe. The pilots are serving customers with physical and virtual cybernetic avatars}
 \label{fig:teaser}
\end{teaserfigure}

\maketitle

\section{Introduction}

The benefits of telepresence robots go beyond the ability for people to move around instantly. They have the potential to extend a person's physical capabilities; AR and VR avatars can further extend their abilities and even move beyond their real physical limits. This is very attractive and important, especially for people with disabilities.

Our mission is to solve loneliness through technology. We have developed an avatar robot system that enables people who have given up communication due to disabilities to work from home, where an operator (we call him or her "pilot") selectively switches between two types of avatar robots via the Internet \cite{Takeuchi20}. We also prepare and an Avatar robot cafe the avatar work, which is already in operation in Japan (Figure 1: (a)) \cite{DAWN21}. The pilot controls the robot from his/her home by operating an application on a tablet or PC supported according to his/her disability (Figure 2).

\section{Meta Avatar Robot Cafe}

In addition to avatar robots, the Meta Avatar robot cafe offers AR avatars and VR avatars. The actual control screen seen by the pilot is shown in Figure 1 (b) and (c). A digital twin of the cafe is also available for VR avatars.

One of the important aspects of telepresence robots is to share experiences with others in real world. Sharing the experience in real world with others makes the remote location feel real and in the moment. With a physical avatar robot, the pilot can feel that he or she is in the cafe for real through serving customers there by remote control. This is a great advantage for people with disabilities who have difficulty going out. On the other hand, the robot is subject to physical limitations on its movements. For safety reasons, robots are restricted in their active movements. To solve this limitation problem, we have prepared an AR avatar. AR and VR applications can extend physical capabilities beyond robots. Flashy movements like dancing and kicking, which are never possible with their real body, are popular motions for the pilots.

\begin{figure}[h]
  \centering
  \includegraphics[width=\linewidth]{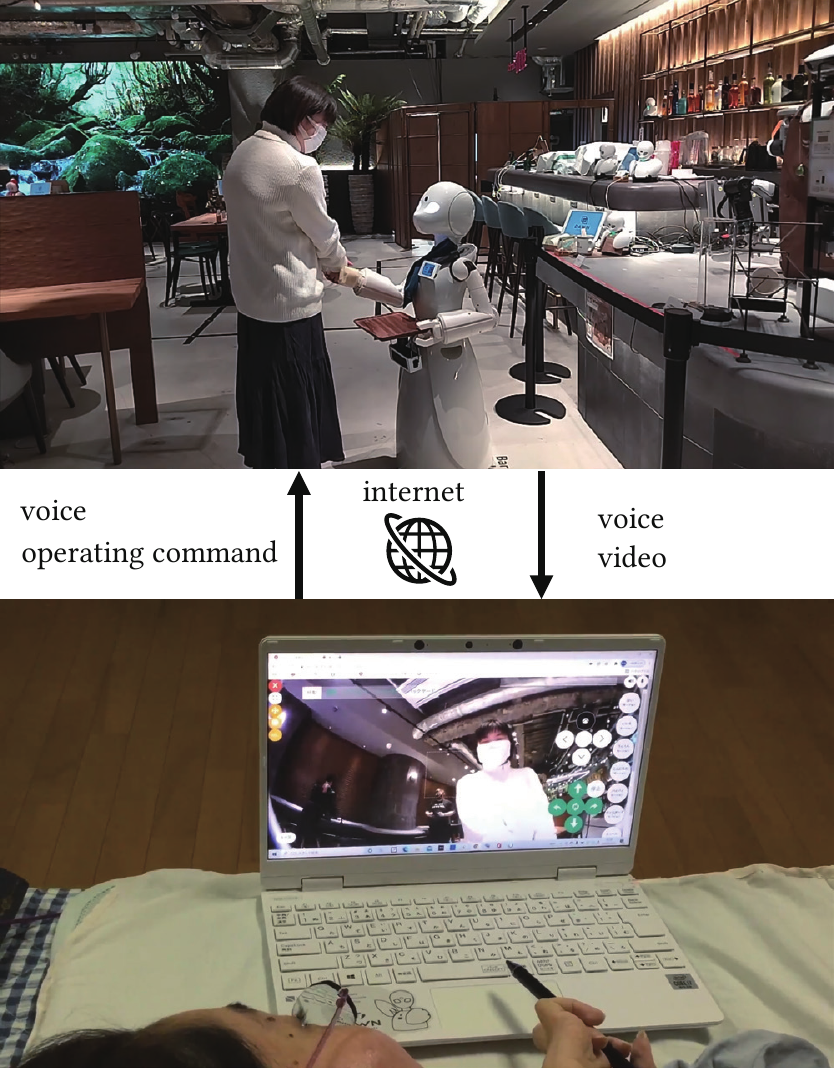}
  \caption{Operation of the avatar robot OriHime-D by a pilot with disabilities at the Avatar robot cafe.}
  \Description{A pilot with disability operates OriHime-D at the cafe, from her home.}
\end{figure}

\section{Demonstration at the avatar robot cafe DAWN ver.$\beta$}

The most important thing to remember about the Meta Avatar robot cafe is that AR, VR and the real space cafe mean exactly the same thing to a pilot operating from a remote location. This may seem obvious, but it is a very important realization. For the pilot, there is no sensory difference between the physical robot body and the AR avatar body, as long as they can operate from the same screen. They can freely change and manipulate the body according to what they want to do.

As one example, a bedridden person with a disability said that he sat in a chair for the first time in several years after controlling his avatar. This kind of experience is difficult to achieve with telepresence robots alone. This experience could only have been realized by using a physical robot to interact with a person and gain a sense of reality at a distance, and then switching to an AR or VR avatar free from movement restrictions.

\section{Conclusions}

In addition to the telepresence robot, the Meta Avatar robot cafe has a VR space that is connected to reality, and a social system of working at the cafe all in one set. That is why we were able to realize this kind of experience by people with disabilities. We believe this is the major value of the Meta Avatar robot cafe. We believe that if we can provide the technology to switch between various physical and virtual  avatars with different characteristics and their social environment, we can realize a society in which people can participate in various social activities even if they become bedridden.

\begin{acks}
  This work was supported by JST Moonshot R\&D Program “Cybernetic being” Project (Grant number JPMJMS2013).
\end{acks}

\bibliographystyle{ACM-Reference-Format}
\bibliography{siggraph22emergingtechnologies-13}

\end{document}